\documentclass[final,5p,times]{elsarticle}

\usepackage{graphicx}
\usepackage{booktabs}
\usepackage{bm}
\usepackage{array}
\usepackage{hyperref}
\usepackage{tabularx,dcolumn}
\usepackage{amsmath,amsfonts,amssymb}

\biboptions{sort&compress}

\begin{document}

\begin{frontmatter}
  \title{First-order corrections to semiclassical Gaussian partition functions
    for clusters of atoms}

  \author{Holger Cartarius}
  \ead{Holger.Cartarius@weizmann.ac.il}
  \author{Eli Pollak}
  \address{Chemical Physics Department, Weizmann Institute of Science,
    76100 Rehovot, Israel}
  \date{\today}

  \begin{abstract}
    Gaussian approximations to the Boltzmann operator have proven themselves in
    recent years as useful tools for the study of the thermodynamic properties
    of rare gas clusters. They are, however, not necessarily correct at very
    low temperatures. In this article we introduce a first-order correction
    term to the frozen Gaussian imaginary time propagator and apply it to the
    argon trimer. Our findings show that the correction term provides objective
    access to the quality of the propagator's results and clearly defines the
    ``best'' Gaussian width parameter. The strength of the correction monitored
    as a function of the temperature indicates that the results of the Gaussian
    propagator become questionable below a certain temperature. The interesting
    thermodynamic transition from a bounded trimer to three body dissociation
    lies in the temperature range for which the Gaussian approximation is
    predicted to be accurate.
  \end{abstract}

\end{frontmatter}

\section{Introduction}

The imaginary time or Boltzmann operator $\exp(-\beta \hat{H})$ belongs to
the most important quantities needed for understanding the equilibrium
properties of multi-dimensional systems at finite temperatures. The
derivatives of its trace give access to the mean energy and specific heat,
which allow for the investigation of thermodynamic properties of atomic clusters
\cite{Neirotti00a,Predescu03a,Frantsuzov04a,Predescu05a,White05a,%
  Frantsuzov06a,Perez10a}. For systems with many degrees of freedom it is,
however, still a challenge to evaluate the Boltzmann operator although it is
accessible with Monte Carlo methods \cite{Berne86a,Makri99a,Ceperley03a},
since the necessary integrals can become numerically expensive, in particular
at low temperatures. Approximations are essential to evaluate the Boltzmann
operator for clusters of a few dozen atoms.

Semiclassical initial value representations propagating Gaussian wave packets
of the form
\begin{multline}
  \langle \bm{x} | g \rangle =
   \left (\pi^{3N} |\det \bm{G}(\tau)| \right )^{-1/4} \\ \times
   \exp \left ( -\frac{1}{2} [\bm{x}-\bm{q}(\tau)]^\mathrm{T}
    \bm{G}(\tau)^{-1} [\bm{x}-\bm{q}(\tau)] + \gamma(\tau) \right ) ,
  \label{eq:intro_gaussian}
\end{multline}
for the Boltzmann operator have been developed
\cite{Frantsuzov03a,Frantsuzov04a,Zhang09a} and successfully applied to the
study of thermodynamic properties of clusters
\cite{Frantsuzov03a,Frantsuzov04a,Predescu05a,Frantsuzov06a,Frantsuzov08a,%
  Cartarius11a}. In particular the time-evolved Gaussian approximation
developed by Mandelshtam and co-workers \cite{Frantsuzov03a,Frantsuzov04a}
has become an important tool which has been applied to thermodynamic
properties of a large number of clusters
\cite{Frantsuzov03a,Frantsuzov04a,Predescu05a,Frantsuzov06a,Frantsuzov08a}.
This method is based on so-called \emph{thawed} Gaussian wave packets, where
the matrix of Gaussian width parameters $\bm{G}(\tau)$ changes with (imaginary)
time $\tau$. Recently a \emph{frozen} Gaussian propagator with a
constant width matrix $\bm{G}(\tau) = \bm{G}(0)$ suggested by Zhang et al.\
\cite{Zhang09a} was shown to provide results competitive with the time-evolved
Gaussian approximation using a single particle ansatz
\cite{Cartarius11a} to simplify the computation of the time-dependent width
matrix. Since no equations of motion for the $N\times N$ elements
of $\bm{G}$ have to be propagated, the frozen Gaussian method is numerically
much cheaper.

Despite the large success and broad applicability of Gaussian methods for
the evaluation of the mean energy and the specific heat of clusters
one has to keep in mind that they are semiclassical approximations. As pointed
out by Liu and Miller \cite{Liu06a} they are not expected to be accurate at
very low temperatures where quantum effects are large. In the simplest case
one observes a shift of the ground state energy away from the exact result
\cite{Cartarius11a}. The low-temperature range is, however, very often the
most important and crucial for the thermodynamic properties of rare gas
cluster. For example, Gaussian based computations on clusters of light atoms,
e.g., $\mathrm{Ne}_{13}$ and $\mathrm{Ne}_{38}$
\cite{Predescu05a,Frantsuzov06a}, predict novel low temperature quantum
effects such as liquid-like zero temperature structures of $\mathrm{Ne}_{38}$
as compared to a solid-like structure predicted from classical mechanics
\cite{Frantz92a}. These predictions are based on the Gaussian approximation
\cite{Frantsuzov03a,Frantsuzov04a,Liu06a,Frantsuzov06a,Frantsuzov08a}
and have to be verified.

A systematic approach connecting the Gaussian semiclassical initial value
approximations with exact quantum mechanics is the generalized
time-de\-pen\-dent perturbation approach developed by Pollak and coworkers
\cite{Shao06a,Zhang09a}.
Within this framework the Gaussian approximations can be considered as a
lowest-order approximation of a series converging to the exact quantum
Boltzmann operator. It has been shown that both the thawed and frozen Gaussian
versions of this series converge rapidly to the numerically exact answer for
a one-dimensional double well potential \cite{Zhang09a,Conte10a}. In this
paper we demonstrate that the corrections to the Gaussian imaginary time
propagator are also practically applicable to higher-dimensional systems, in
particular to atomic clusters. To do so, we apply the frozen Gaussian series
to the partition function of the argon trimer and calculate its first-order
correction. We show that the correction helps to estimate the quality of the
results and provides objective access to the validity of the Gaussian
approximations. With the first-order terms it is possible to identify a
border temperature below which the Gaussian results become questionable. This
result clearly states that the dissociation process \cite{Perez10a}
discussed earlier with Gaussian approximations \cite{Cartarius11a} is correctly
described by the frozen Gaussian imaginary time propagator since it appears
in the allowed temperature range.

The Gaussian width of the frozen Gaussian propagator is a free parameter but
has an important impact on the accuracy of the physical quantities calculated
with the approximation. It is demonstrated in this article that the first-order
correction provides an unambiguous method for optimizing the value which is
based on minimizing the ratio between the first-order correction to the
zeroth-order term for the Boltzmann operator. This choice improves the
results for the mean energy and the specific heat. Including the first-order
correction they reach the same quality as a fully-coupled thawed Gaussian
computation over a wide range of temperatures.

This article is organized as follows. In Sec.\ \ref{sec:propagator} we review
the most important parts of the frozen Gaussian series representation of the
Boltzmann operator and introduce the first-order correction to the partition
function for clusters of atoms. The theory is then applied to the argon trimer
in Sec.\ \ref{sec:argon_trimer}. We introduce the system (Sec.\
\ref{sec:argon_system}), show how the first-order correction term may be used
to determine the best width parameter (Sec.\ \ref{sec:choice_width}), consider
the influence of an artificial confinement on the correction (Sec.\
\ref{sec:confinement}) and present the first-order corrected mean energy
and specific heat for the dissociation process of the cluster (Sec.\
\ref{sec:dissociation}), which allow for a clear statement on the quality
of the Gaussian approximation. The required numerical effort is analyzed in
Sec.\ \ref{sec:numerical}. Conclusions are drawn in Sec.\ \ref{sec:conclusion}.

\section{Frozen Gaussian approximation to the partition function
  and first-order corrections}
\label{sec:propagator}

\subsection{Frozen Gaussian series representation}

Zhang et al.\ \cite{Zhang09a} proposed a frozen Gaussian form of the imaginary
time propagator $K(\beta) = \exp(-\beta \hat{H})$ at inverse temperature $\beta
= 1/(\mathrm{k}T)$ and developed its higher-order corrections in the framework
of the generalized time-dependent perturbation series
\cite{Zhang03a,Pollak03a,Shao06a}. The frozen Gaussian coherent state has the
form
\begin{multline}
  \langle \bm{x} | g(\bm{p}(\tau),\bm{q}(\tau),\bm{\Gamma}) \rangle 
  = \left ( \frac{\det(\bm{\Gamma})}{\pi^{3N}} \right )^{1/4} \\
  \times \exp \Biggl ( -\frac{1}{2} [\bm{x} - \bm{q}(\tau)]^\mathrm{T}
  \bm{\Gamma} [\bm{x} - \bm{q}(\tau)] + \frac{i}{\hbar} \bm{p}^\mathrm{T}(\tau) 
  \cdot [\bm{x}-\bm{q}(\tau)] \Biggr ) ,
\end{multline}
with the symmetric positive definite matrix of constant width parameters
$\bm{\Gamma}$ and the dynamical variables $\bm{q}(\tau)$ and $\bm{p}(\tau)$.
Approximating a solution to the Bloch equation
\begin{equation}
  -\frac{\partial}{\partial \tau} |\bm{q}_0,\tau \rangle
  = H  |\bm{q}_0,\tau \rangle
 \label{eq:bloch}
\end{equation}
by propagating a frozen Gaussian wave packet, one finds that the imaginary
time propagator matrix element has the form
\begin{multline}
  \langle \bm{x}' | K_0(\tau) | \bm{x} \rangle
  = \det(\bm{\Gamma}) \exp \left ( -\frac{\hbar^2}{4} \mathrm{Tr}(\bm{\Gamma})
    \tau \right ) \\ \times 
  \sqrt{\det \left ( 2 \left [ \bm{1} - \exp (-\hbar^2
        \bm{\Gamma} \tau) \right ]^{-1} \right )} \\
  \times \exp \Biggl ( -\frac{1}{4} [\bm{x}' - \bm{x}]^\mathrm{T} \bm{\Gamma}
    [\tanh(\hbar^2 \bm{\Gamma} \tau/2)]^{-1} [\bm{x}'-\bm{x}] \Biggr ) \\
  \times \int \frac{d\bm{q}^{3N}}{(2\pi)^{3N}}
  \exp \Biggl (-2 \int_0^{\tau/2} d\tau' \langle V(\bm{q}(\tau')) \rangle \\
  - [\bm{\bar{x}}-\bm{q}(\tau/2)]^\mathrm{T} \bm{\Gamma}
  [\bm{\bar{x}}-\bm{q}(\tau/2)] \Biggr )
  \label{eq:prop0}
\end{multline}
with $\bm{\bar{x}} = (\bm{x}' + \bm{x})/2$. The only remaining dynamical
variables in this representation of the propagator are the components of the
vector $\bm{q}(\tau)$, which follow the equations of motion
\begin{equation}
  \frac{\partial \bm{q}(\tau)}{\partial \tau} = -\bm{\Gamma}^{-1}
  \langle \nabla V(\bm{q}(\tau)) \rangle .
\end{equation}
The angle brackets symbolize Gaussian averaged quantities of the form
\begin{multline}
  \langle h(\bm{q}) \rangle =  \left ( \frac{\det(\bm{\Gamma})}{\pi^{3N}}
  \right )^{1/2} \\
  \times \int_{-\infty}^{\infty} d\bm{x}^{3N} \exp \left ( - [\bm{x} 
    - \bm{q}]^\mathrm{T} \bm{\Gamma} [\bm{x} - \bm{q}] \right ) h(\bm{x}).
  \label{eq:Gaverage}
\end{multline}

In the framework of the generalized time-dependent perturbation theory,
$K_0(\tau)$ is the zeroth-order approximation to the exact Boltzmann operator.
Since the Bloch equation \eqref{eq:bloch} is not solved exactly by the
propagator \eqref{eq:prop0}, one can define the correction operator
\cite{Shao06a}
\begin{equation}
  C(\tau) = - \frac{\partial}{\partial \tau} K_0(\tau) - H K_0(\tau) .
  \label{eq:corrop}
\end{equation}
The formal solution of Eq.\ \eqref{eq:corrop} is
\begin{equation}
  K_0(\tau) = K(\tau) - \int_0^\tau d\tau' K(\tau-\tau') C(\tau') .
\end{equation}
With the assumption that the exact propagator $K(\tau)$ can be expressed
in terms of a series
\begin{equation}
  K(\tau) = \sum_{n=0}^{\infty} K_j(\tau) ,
\end{equation}
where $K_j(\tau) \sim C(\tau)^j$ are terms with ascending power in the
small correction, one can write down the recursion relation
\begin{equation}
  K_{j+1}(\tau) = \int_{0}^{\tau} d\tau' K_{j}(\tau-\tau') C(\tau')
  \label{eq:recursion}
\end{equation}
for the calculation of higher-order terms.

\subsection{First-order corrections to the partition function}

In the framework of the series expansion the lowest-order approximation
to the quantum partition function is simply the trace of the zeroth-order
propagator \cite{Zhang09a},
\begin{multline}
  Z_0(\beta) = \mathrm{Tr} \left [ K_0(\beta) \right ]
  = \sqrt{\det(\bm{\Gamma})} \exp \left ( -\frac{\hbar^2}{4}
    \mathrm{Tr}(\bm{\Gamma}) \beta \right ) \\
  \times \sqrt{\det \left ( \left [ \bm{1} - \exp (-\hbar^2 \bm{\Gamma}
        \beta) \right ]^{-1} \right )} \\
  \times \int_{-\infty}^\infty \frac{d\bm{q}^{3N}}{(2\pi)^{N/2}}
  \exp \left (-2 \int_0^{\beta/2} d\tau \langle V(\bm{q}(\tau)) \rangle
  \right ) .
  \label{eq:pf_FG0}
\end{multline}
Its first-order corrected counterpart
\begin{subequations}
  \begin{equation}
    Z_1(\beta) = Z_0(\beta) + Z_{\mathrm{C}1}(\beta)
  \end{equation}
  is obtained in a similar way. The first-order correction to the propagator
  is evaluated according to Eq.\ \eqref{eq:recursion} and then the correction
  term for the partition function is
  \begin{multline}
    Z_{\mathrm{C}1}(\beta) = \mathrm{Tr} \left [ K_1(\beta) \right ]
    \\ = \int_0^\beta d\tau \int d\bm{x}'^{3N} \int d\bm{x}^{3N}
    \langle \bm{x}' | K_0(\beta-\tau) | \bm{x} \rangle
    \langle \bm{x} | C(\tau) | \bm{x}' \rangle .
    \label{eq:pf_corr1}
  \end{multline}
\end{subequations}
More details of the correction operator and the first-order term to the
partition function can be found in \ref{app:correction_terms}. In
particular it is shown, how the $\bm{x}'$ and $\bm{x}$ integrations in
Eq.\ \eqref{eq:pf_corr1} can be performed analytically if the potential of
the system can be expressed in terms of Gaussians. This makes the evaluation
of the correction terms efficient.

Note that the first-order correction term \eqref{eq:pf_corr1} is not
symmetric. Here we used the ``left'' correction operator defined by Zhang
et al.\ \cite{Zhang09a}. This has, however, no influence on the partition
function due to the trace in Eq.\ \eqref{eq:pf_corr1}.

\section{First-order calculations for the argon trimer}
\label{sec:argon_trimer}

\subsection{Potential and confinement}
\label{sec:argon_system}

To be consistent with previous studies of the system
\cite{Perez10a,Gonzales-Lezana99a,Cartarius11a} we model the pairwise
interaction of the argon atoms by a Morse potential
\begin{equation}
  V(r_{ij}) = D \left ( \exp \left [ -2\alpha (r_{ij}-R_\mathrm{e}) \right ]
    - 2 \exp \left [ -\alpha (r_{ij}-R_\mathrm{e}) \right ]  \right)
  \label{eq:Morse_potential}
\end{equation}
with the distance $r_{ij}$ between particles $i$ and $j$ and the Morse
parameters $D = 99.00\,\mathrm{cm}^{-1}$, $\alpha = 1.717\,\text{\r{A}}$,
and $R_\mathrm{e} = 3.757\,\text{\r{A}}$. Since the Gaussian averaged quantities
according to Eq.\ \eqref{eq:Gaverage} and the $\bm{x}'$ and $\bm{x}$
integrations in the first-order correction term \eqref{eq:pf_corr1} can be
performed analytically if the potential can be expressed in terms of
Gaussians, we fitted the Morse potential \eqref{eq:Morse_potential} to a sum
of three Gaussians
\begin{equation}
  V(|\bm{r}_i - \bm{r}_j|) = \sum_{p=1}^{3} c_p e^{-\alpha_p r_{ij}^2} ,
  \qquad r_{ij} = |\bm{r}_i - \bm{r}_j| ,
  \label{eq:Gaussian_fit}
\end{equation}
with the parameters listed in Table \ref{tab:Gaussian_parameters}.
\begin{table}
  \caption{\label{tab:Gaussian_parameters}Parameters used in the Gaussian
    representation \eqref{eq:Gaussian_fit} of the potential, taken from
    Ref.\ \cite{Cartarius11a}.}
  \begin{tabularx}{\columnwidth}{XD{.}{.}{8}D{.}{.}{5}X}
    \toprule
    $p$ & \multicolumn{1}{c}{$c_p$ [$\mathrm{cm}^{-1}$]} &
    \multicolumn{1}{c}{$\alpha_p$ [$\text{\r{A}}^{-2}$]} & \\
    \midrule
    1 & 3.296\times 10^{5} & 0.6551 & \\
    2 & -1.279\times 10^{3} & 0.1616 & \\
    3 & -9.946\times 10^{3} & 6.0600 & \\
    \bottomrule
  \end{tabularx}
\end{table}

Since all $\bm{x}'$ and $\bm{x}$ integrations in the zeroth-order approximation
and the first-order correction term of the propagator can be performed
analytically with the Gaussian potential \eqref{eq:Gaussian_fit} only the
integration in the $\bm{q}$ space and the imaginary time integration in Eq.\
\eqref{eq:pf_corr1} remain for numerical computation. The $\tau$ integration
is one-dimensional and can be done with a standard integrator. The
multi-dimensional Monte Carlo sampling in the $\bm{q}$ variables is done with
a standard Metropolis algorithm as outlined in Ref.\ \cite{Frantsuzov04a}.

To converge the integrals the $\bm{q}$ space is restricted by the condition
$|\bm{q} - \bm{R}_\mathrm{cm}| < R_\mathrm{c}$, where $R_\mathrm{c}$ is a confining
radius and $\bm{R}_\mathrm{cm}$ is the center of mass of the cluster. That is,
no atom may leave the center of mass beyond a certain distance $R_\mathrm{c}$.
It is also possible to introduce the confinement via an additional steep
potential \cite{Predescu03a} as
\begin{equation}
  V_\mathrm{c}(\bm{r}) \propto \sum_{i=1}^{N} \left ( \frac{\bm{r}_i
      - \bm{R}_\mathrm{cm}}{R_\mathrm{c}} \right )^{20} .
  \label{eq:confinement}
\end{equation}
As has been shown previously \cite{Predescu03a,Etters75a,Cartarius11a}, the
value of $R_\mathrm{c}$ has a critical influence on the thermodynamic
properties. For the first-order correction term \eqref{eq:pf_corr1} two
$\bm{q}$ integrations, viz.\ in the propagator and in the correction operator,
have to be taken into account. Here, the restriction on $\bm{q}$ is applied
for the propagator part \eqref{eq:prop0}. Due to the coupling via $\bm{x}'$
and $\bm{x}$ the restriction automatically affects also the $\bm{q}$
variable in the correction operator. More details can be found in
\ref{app:correction_terms}.

In this article we concentrate on a confinement of $R_\mathrm{c} =
10\,\text{\r{A}}$. It is known that this confinement does not describe the
dissociation process fully since it is too restrictive. However, the
restricted cluster shows a very broad transition from the bounded system to
three free particles, and thus allows for a clear analysis of the
correction term's influence on the results around the dissociation, which is
the interesting physical effect.

\subsection{Choice of the width parameter}
\label{sec:choice_width}

In the frozen Gaussian ansatz the matrix of width parameters $\bm{\Gamma}$ is
constant in imaginary time $\tau$, however,
the choice of $\bm{\Gamma}$ has a critical influence on the quality of the
results \cite{Zhang09a}. In a previous study \cite{Cartarius11a} we found
that the width matrix
\begin{equation}
  \bm{\Gamma} = \begin{pmatrix}
    (\bm{D}_1+2\bm{D}_2)/3 & (\bm{D}_1-\bm{D}_2)/3 & (\bm{D}_1-\bm{D}_2)/3 \\
    (\bm{D}_1-\bm{D}_2)/3 & (\bm{D}_1+2\bm{D}_2)/3 & (\bm{D}_1-\bm{D}_2)/3 \\
    (\bm{D}_1-\bm{D}_2)/3 & (\bm{D}_1-\bm{D}_2)/3 & (\bm{D}_1+2\bm{D}_2)/3 \\
  \end{pmatrix}
\end{equation}
with two $3\times 3$-submatrices $\bm{D}_1 = D_1 \bm{1}$ and $\bm{D}_2 = D_2
\bm{1}$, where $\bm{1}$ is the $3\times 3$ identity matrix, provides results
competitive with thawed Gaussian calculations using a single-particle ansatz.
The parameter $D_1$ represents the free center of mass motion and should be
chosen as small as possible. The Gaussian width for the relative coordinates
is represented by $D_2$. In Ref.\ \cite{Cartarius11a} an optimum
value of $D_2 = 25\,\text{\r{A}}^{-2}$ was found. It yielded the best
approximation to a numerically exact path integral Monte Carlo calculation
\cite{Perez10a} and thawed Gaussian approximations
\cite{Cartarius11a}. In the same work it was found that all choices of
$D_1$ below a certain value lead to the same mean energies and that $D_1 =
0.1\,\text{\r{A}}^{-2}$ belongs to this range in which $D_1$ is small enough.
These parameters also provide the best and lowest approximation to the ground
state energy. Thus, the minimum of the ground state energy could also be used
to search for the optimum value of the width parameters. It provides, however,
no access to the quality of the results if the exact ground state energy is
unknown.

A systematic, objective and internally consistent approach to the determination
of the optimum width parameter is made possible by considering the first-order
correction. The definition of the correction operator in Eq.\ \eqref{eq:corrop}
makes clear that it measures the deviation of the zeroth-order term from the
exact result. One should thus minimize the ratio of the first-order
[Eq.\ \eqref{eq:pf_corr1}] and zeroth-order [Eq.\ \eqref{eq:pf_FG0}]
contributions to the partition function by varying the width parameter matrix
and thus find the optimum value.

\begin{figure}[tb]
  \includegraphics[width=\columnwidth]{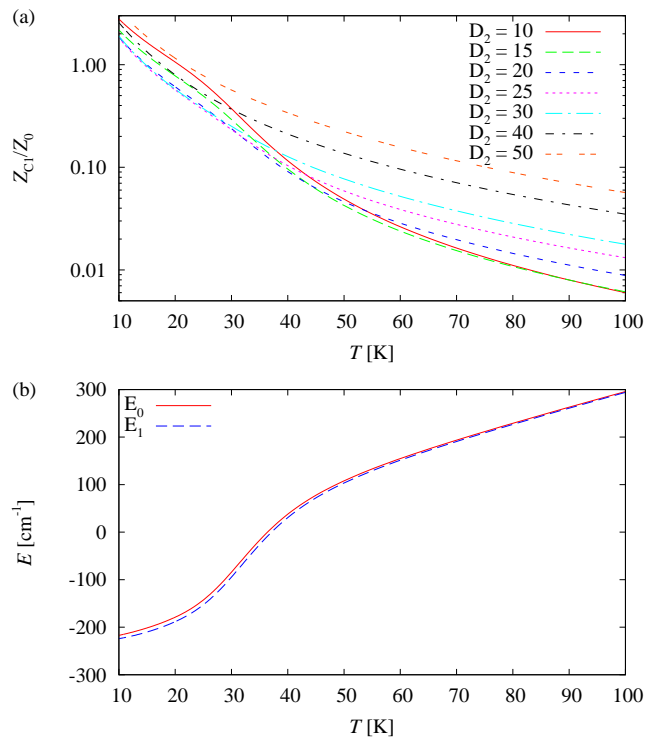}
  \caption{\label{fig:parameterD2} (a) The ratio of the contribution of the
    first- and zeroth-order terms in the series expansion of the partition
    function is plotted as a function of the temperature $T$ for varying values
    of the width parameter $D_2$ (given in units of $\text{\r{A}}^{-2}$) and
    $D_1 = 0.1\,\text{\r{A}}^{-2}$. For the bounded system ($T \lessapprox
    30\,\mathrm{K}$) the range $D_2 = 20\dots 30\,\text{\r{A}}^{-2}$ leads to
    the minimal ratio whereas for temperatures above the transition temperature
    smaller values of $D_2$ lead to smaller relative corrections. (b) The
    zeroth-order approximation to the mean energy $E_0$ and its first-order
    corrected counterpart $E_1$ is shown to visualize the transition from the
    bounded to the unbounded system for $D_1 = 0.1\,\text{\r{A}}^{-2}$ and
    $D_2 = 25\,\text{\r{A}}^{-2}$.}
\end{figure}

In Fig.\ \ref{fig:parameterD2}(a)
we plot the ratio $Z_\mathrm{C1}/Z_0$ for several values of the parameters $D_2$
and $D_1 = 0.1\,\text{\r{A}}^{-2}$.
The data was calculated for a cluster with a confinement of $R_\mathrm{c} =
10\,\text{\r{A}}$. The zeroth- and first-order mean energies in
Fig.\ \ref{fig:parameterD2}(b) visualize the transition from a bounded to an
unbounded cluster. For temperatures at which the system is bounded, i.e., $T
\lessapprox 30\,\mathrm{K}$ an optimum range for the width parameter at which
the contribution of the correction term $Z_{\mathrm{C}1}$ is minimal can be
found. Below the transition temperature the lines for $D_2$ between $20\,
\text{\r{A}}^{-2}$ and $30\,\text{\r{A}}^{-2}$ are close to each other, the
smallest value is found for $D_2 = 25\, \text{\r{A}}^{-2}$. This confirms our
previous zeroth-order study of the argon trimer \cite{Cartarius11a}, where the
same value of $D_2 = 25\,\text{\r{A}}^{-2}$ provided the best agreement with
numerically exact path integral Monte Carlo methods. At higher temperatures,
where the cluster passes through a transition to three almost free atoms, the
situation changes. The ratio becomes smaller as the magnitude of $D_2$
decreases. In the limit of three free particles, the frozen Gaussian
approximation is exact, in the limit that the width matrix vanishes. One thus
expects that in this limit, the smaller values of the width parameters would
provide a better approximation.

\subsection{Influence of the confining sphere}
\label{sec:confinement}

\begin{figure}[tb]
  \includegraphics[width=\columnwidth]{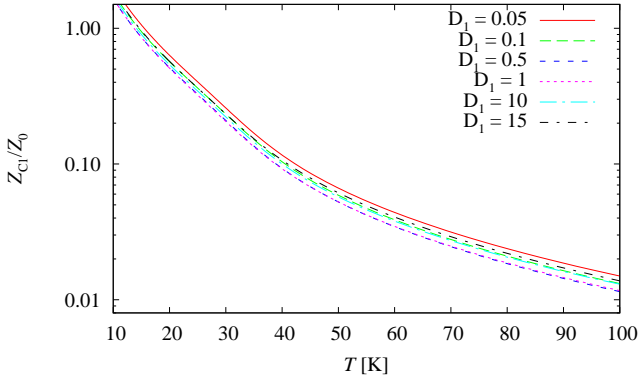}
  \caption{\label{fig:parameterD1} The ratio of the contribution of the first-
    and zeroth-order terms in the series expansion of the partition function
    is plotted as a function of the temperature $T$ for varying values of the
    width parameter $D_1$ (given in units of $\text{\r{A}}^{-2}$) and $D_2
    = 25\,\text{\r{A}}^{-2}$. The minimal ratio is found in the range $D_1 =
    0.5 \dots 1\,\text{\r{A}}^{-2}$. The cluster is limited by a confinement
    with $R_\mathrm{c} = 10\,\text{\r{A}}$.}
\end{figure}

As already mentioned, for free particles, the frozen Gaussian propagator
yields the exact partition function in the limit $\bm{\Gamma} \to \bm{0}$.
Thus, one expects that the free center of mass motion, described by $D_1$,
demands $D_1$ to be as small as possible. In practice one observes, however,
that $D_1$ shows a minimum correction at a finite value. This is a result of
the confinement of the atoms, which has the same effect as adding a confining
potential \eqref{eq:confinement}, i.e., the particles are not really free.
The effect can be observed in Fig.\ \ref{fig:parameterD1},
where the ratio $Z_\mathrm{C1}/Z_0$ is plotted for several choices of
$D_1$ and $D_2 = 25\,\text{\r{A}}^{-2}$. As can be seen the minimum correction
is achieved for the finite values $D_1 = 0.5 \dots 1\,\text{\r{A}}^{-2}$ if a
confinement of $R_\mathrm{c} = 10\,\text{\r{A}}$ is used.

The influence of the confinement can also be found in the study of the width
parameter $D_2$ for the internal degrees of freedom. For high temperatures,
i.e., in the limit of three free particles a smaller value of $D_2$ should
provide a better approximation. Thus, one expects always a smaller contribution
$Z_\mathrm{C1}$ for smaller values of $D_2$. However, the strengths of the
first-order corrections for $D_2 = 10\, \text{\r{A}}^{-2}$ and $D_2 =
15\,\text{\r{A}}^{-2}$ are almost the same in the high-temperature limit, even
beyond the temperatures shown in the figure.

\subsection{Correction at low temperatures and the first-order corrected
  mean energy}
\label{sec:dissociation}

\begin{figure}[tb]
  \includegraphics[width=\columnwidth]{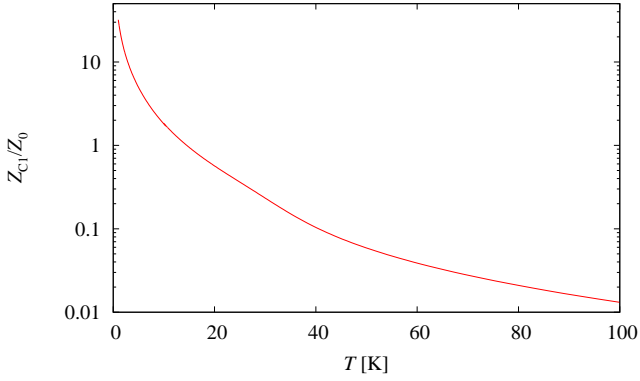}
  \caption{\label{fig:low_temperature} Relative strength of the first-order
    correction to the partition function at lower temperatures for $R_\mathrm{c}
    = 10\,\text{\r{A}}$, $D_1 = 0.1\,\text{\r{A}}^{-2}$, and $D_2 =
    25\,\text{\r{A}}^{-2}$. It diverges in the limit $T \to 0$.}
\end{figure}

Figures \ref{fig:parameterD2}(a) and \ref{fig:parameterD1} already indicate
that the correction term increases in importance as the temperature is lowered.
This is not surprising since an imaginary time propagation is performed
approximately and the difference between the approximate and exact solutions
is expected to increase with (imaginary) time. The behavior becomes even
clearer in Fig.\ \ref{fig:low_temperature},
where the relative strength of the first-order correction $Z_\mathrm{C1}/Z_0$
is plotted for the argon trimer with a confinement $R_\mathrm{c} = 10\,
\text{\r{A}}$ and width parameters $D_1 = 0.1\,\text{\r{A}}^{-2}$,
$D_2 = 25\,\text{\r{A}}^{-2}$ down to a temperature of $T = 1\,
\mathrm{K}$. It increases drastically and shows a diverging behavior
in the limit $T \to 0$, i.e., $\beta \to \infty$. The increasing importance of
the first-order correction term at low temperatures is consistent with our
previous zeroth-order study of the system, where we found that the accuracy of
the frozen Gaussian approximation gets worse as $T \to 0$ and the cluster
approaches the ground state \cite{Cartarius11a}. At low temperatures, quantum
effects are strong and, in particular, the Gaussian (zeroth-order)
approximations cannot reproduce the exact ground state energy $E_\mathrm{GS}$
of the cluster. The Gaussian form imposed on the wave function is too severe
and the ground state energy obtained from the zeroth-order approximation is
found to be slightly higher than the exact ground state energy ($E_0 =
E_\mathrm{GS} + \Delta E$ and $\Delta E > 0$).

To further understand the divergence of the first-order term, we may assume
that for large $\beta$ only the ground state contributes to the partition
function and that the exact partition function $Z$ and its zeroth-order
approximated counterpart $Z_0$ behave as
\begin{equation}
  Z \approx e^{-\beta E_\mathrm{GS}} , \qquad Z_0 \approx e^{-\beta (E_\mathrm{GS} +
    \Delta E)} .
\end{equation}
Thus, a first-order correction which is valid also for large $\beta$ would be
expected to diverge, since $Z/Z_0 \approx e^{\beta \Delta E}$.

\begin{figure}[tb]
  \includegraphics[width=\columnwidth]{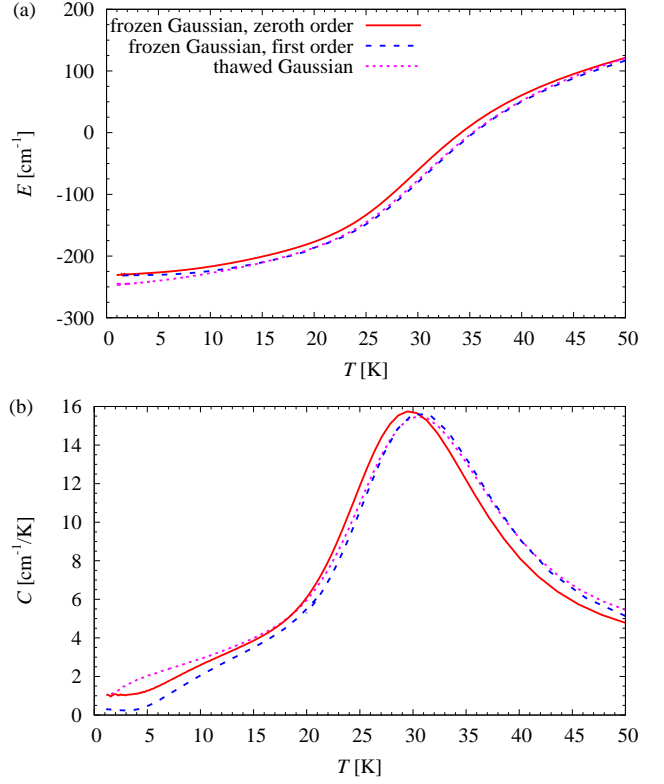}
  \caption{\label{fig:derivates_r10} (a) Temperature dependence of the
    mean energy of the argon trimer for $R_\mathrm{c} = 10\,\text{\r{A}}$,
    $D_1 = 0.1\,\text{\r{A}}^{-2}$, and $D_2 = 25\,\text{\r{A}}^{-2}$. Shown are
    the zeroth-order (solid line) and first-order (dashed line) frozen
    Gaussian results, and the fully-coupled thawed Gaussian approximation
    (dotted line). (b) Specific heat for the same parameters.
    Both calculations indicate a substantial improvement due to the first-order
    correction in the temperature region of the transition and above. For
    lower temperatures the correction breaks down, indicating that the frozen
    Gaussian approximation is no longer sufficiently accurate.}
\end{figure}

More interesting is, however, the influence of the first-order correction on
the derivatives of the partition function, which are required for the
physically meaningful average energy
\begin{subequations}
  \begin{equation}
    E = \mathrm{k} T^2 \frac{\partial \ln Z}{\partial T}
    \label{eq:def_mean_energy}
  \end{equation}
  and specific heat
  \begin{equation}
    C = \frac{\partial E}{\partial T} .
  \end{equation}
\end{subequations}
Figure \ref{fig:derivates_r10}
shows the temperature dependence of the mean energy and specific heat of the
argon trimer for $R_\mathrm{c} = 10\,\text{\r{A}}$, $D_1 = 0.1\,
\text{\r{A}}^{-2}$, and $D_2 = 25\,\text{\r{A}}^{-2}$, i.e., the same parameters
as in Fig.\ \ref{fig:low_temperature}. The frozen Gaussian results for the
zeroth- and first-order approximations are compared with a fully-coupled thawed
Gaussian approximation which, as discussed elsewhere \cite{Cartarius11a} is
expected to be more accurate then the frozen Gaussian approximation. As one can
see in Fig.\ \ref{fig:derivates_r10}(a) the first-order correction for the
frozen Gaussian approximation brings the resulting estimate closer to the
fully-coupled thawed Gaussian estimate over most of the temperature range
studied. However, for temperatures $T \lessapprox 12\,\mathrm{K}$ the
first-order correction to the energy becomes smaller and vanishes in the limit
that $T \to 0$. This also influences the specific heat as can be seen in Fig.\
\ref{fig:derivates_r10}(b). In the temperature range of the transition from
bounded to unbounded motion, the first-order corrected frozen Gaussian
calculations agree well with the fully-coupled thawed Gaussian counterpart.
For lower temperatures this is no longer so. The divergence is especially
noticeable in the low temperature limit for the specific heat where the
first-order correction significantly increases the difference between the
fully-coupled thawed Gaussian estimate (which is rather accurate as known from
numerically exact computations, see Ref.\ \cite{Cartarius11a}) and the frozen
Gaussian based estimate.

\begin{figure}[tb]
  \includegraphics[width=\columnwidth]{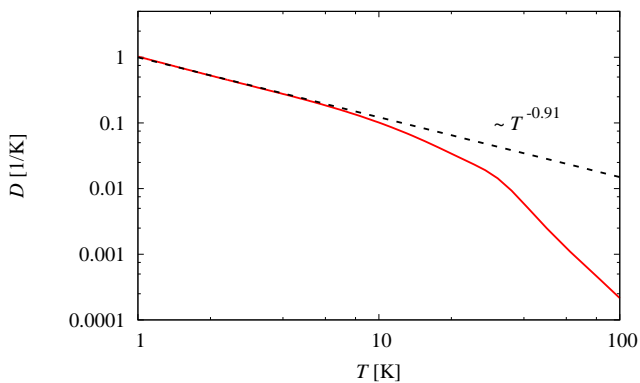}
  \caption{\label{fig:energy_correction} Double logarithmic plot of the
    difference $D$ between the first derivatives of the logarithms of the
    zeroth- and first-order partition functions as defined in
    Eq.\ \eqref{eq:deriv1}. In the limit of low temperatures the divergence
    behaves as $D \sim T^{-0.91}$, which is not strong enough to compensate for
    the $T^2$ term in Eq.\ \eqref{eq:def_mean_energy} needed to obtain a
    nonvanishing energy correction.}
\end{figure}

Interestingly, although the first-order correction term at very low
temperatures is much larger than the zeroth-order frozen Gaussian estimate
(see Fig.\ \ref{fig:low_temperature}), it does not lead to any change in the
estimate of the ground state energy. As may be seen from Fig.\
\ref{fig:derivates_r10}(a) the increase of the correction to
the partition function is not ``fast enough''. This is further demonstrated
in Fig.\ \ref{fig:energy_correction},
where the difference between the first derivatives of the logarithms of the
zeroth-order and first-order partition functions,
\begin{equation}
  D = \frac{\partial \ln Z_0}{\partial T} - \frac{\partial \ln Z_1}{\partial T}
  \label{eq:deriv1}
\end{equation}
is plotted vs. the temperature. From Fig.\ \ref{fig:low_temperature} one
expects this difference to diverge and this is evidently the case.
However, the first-order correction term can lead to a correction for the
estimate of the ground state energy only if the temperature derivative
diverges as $T^2$, as can be seen from Eq.\ \eqref{eq:def_mean_energy}. We
find, however, that below a critical temperature the difference of the
derivatives behaves as $D \sim T^{-0.91}$, and this is not sufficient. As a
result the first-order correction to the ground state energy vanishes as found
in Fig.\ \ref{fig:derivates_r10}.

Although the first-order correction term does not lead to a corrected estimate
for the ground state energy, it does provide information on the quality of the
frozen Gaussian approximation. The ratio $Z_\mathrm{C1}/Z_0$ (cf.\ Fig.\
\ref{fig:low_temperature}) and the temperature behavior of the correction to
the mean energy [cf.\ Figs.\ \ref{fig:derivates_r10} (a) and
\ref{fig:energy_correction}] provide an objective measure for the correctness
of the values as estimated from the frozen Gaussian approximation. Only at a
temperature $T \approx 12\,\mathrm{K}$ and below it, the correction becomes
comparable with the zeroth-order estimate, and thus is no longer small.
Approximately at the same temperature the energy correction starts to get
smaller. Both outcomes indicate that the quality of the approximation is
questionable for lower temperatures. However, the dissociation process
described in Ref.\ \cite{Cartarius11a} appears only at higher
temperatures, indicating that the frozen Gaussian approximation accurately
reflects the quantum transition from a bounded moiety to dissociation into
three free particles.

\section{Numerical effort for evaluating the first-order correction}
\label{sec:numerical}

The results presented for the mean energy and the specific heat in Sec.\
\ref{sec:argon_trimer} showed that the quality of the first-order corrected
frozen Gaussian approximation is similar to the fully-coupled thawed
Gaussian method in the interesting range of temperatures. It is therefore of
interest to compare their computational cost. 

The numerically most expensive part is the imaginary time propagation of
the dynamical variables. Since the positions $\bm{q}$ in the zeroth-order
frozen Gaussian propagator \eqref{eq:prop0} and the correction operator
\eqref{eq:corrop} are the only dynamical variables of a frozen Gaussian
propagator, the number of equations of motion for the first-order corrected
partition function still scales linearly with the particle number. For $N$
particles one has to solve $6 N + 2$ equations of motion, in which the time
integrations of the averaged potential in the exponential of the propagator and
the correction operator [c.f.\ Eq.\ \eqref{eq:prop0} and
\ref{app:correction_terms}] are included. By contrast the number of dynamical
variables of the fully-coupled thawed Gaussian propagator scales always
quadratically with the number of particles due to the symmetric time-dependent
Gaussian width matrix. Counting all dynamical variables of the time-evolved
Gaussian approximation \cite{Frantsuzov03a,Frantsuzov04a} one obtains
$3N(3N+3)/2+1$ equations of motion. Thus, the imaginary time propagation of
the dynamics is cheaper for the first-order frozen Gaussian partition function.

The first-order correction requires, however, also additional numerical effort.
The need for propagating pairs of $\bm{q}$ trajectories for the propagator and
the correction operator in Eq.\ \eqref{eq:pf_corr1} demands a larger number of
sampling points for the Monte Carlo integration in the two sets of positions
$\bm{q}$. Additionally, the time integration in Eq.\ \eqref{eq:pf_corr1} has
to be evaluated. At this point the importance of an analytical evaluation of
the $\bm{x}'$ and $\bm{x}$ integrations in Eq.\ \eqref{eq:pf_corr1} as
described in \ref{app:correction_terms} becomes clear. Owing to this
simplification one can do without an additional $6N$-dimensional Monte
Carlo sampling for the operator product [$\bm{x}$ integration in Eq.\
\eqref{eq:pf_corr1}] and the trace ($\bm{x}'$ integration). This drastically
reduces the numerical costs. Note that a $3N$-dimensional trace integration
is already required for the zeroth-order approximation of both the frozen and
thawed Gaussian partition functions. It can be and is evaluated analytically
in all cases. With the analytical evaluation of the $\bm{x}'$ and $\bm{x}$
integrations there remain $6N$ $\bm{q}$ integrations for the first-order
corrected frozen Gaussian partition function and $3N$ for its zeroth-order
thawed Gaussian counterpart.

Because of the different numbers of required sampling points there is no general
way for estimating the numerical effort. In the case of the argon trimer
the additional effort for the first-order correction outweighs the lower
number of equations of motion. We used $6.5\times 10^7$ sampling points
for the frozen Gaussian with first-order correction and $1.2\times 10^7$
for the thawed Gaussian to obtain converged results down to low temperatures
of about $1\,\mathrm{K}$. On the same architecture (one Tesla C1060 GPU) the
frozen Gaussian first-order calculation resulted in approximately twice the
time as the zeroth-order thawed Gaussian. Due to the better scaling of the
number of dynamical variables (linear instead of quadratic) one may expect
that this relation changes in favor of the first-order frozen Gaussian variant
with increasing particle number. However, as the results in this article
demonstrate, the information the first-order correction provides on the
validity of the Gaussian approximations is at least as important as the
improvement of the numerical estimate of the physical values.

\section{Conclusions and outlook}
\label{sec:conclusion}

In this article we introduced first-order corrections to a frozen Gaussian
approximation \cite{Zhang09a} of the Boltzmann operator with application to
the thermodynamics of atomic clusters. By using a Gaussian fit of the
underlying potential it is possible to evaluate many of the necessary
integrations of the correction terms analytically such that the first-order
correction becomes viable for systems with ``many'' degrees of freedom. We
applied the correction to a study of the thermodynamics of the argon trimer,
whose dissociation process has been investigated recently
\cite{Perez10a,Cartarius11a}.

The highest value of the correction term is to asses the quality of Gaussian
approximations used in the study of thermodynamic properties of
high-dimensional systems
\cite{Neirotti00a,Predescu03a,Frantsuzov04a,Predescu05a,White05a,%
  Frantsuzov06a,Perez10a,Frantsuzov08a,Cartarius11a}. It is known that these
Gaussian approximations are exact in the high-temperature limit but are not
necessarily correct at low temperatures. The present study shows that the
first-order correction indicates a border temperature below which the results
of the approximate propagators become questionable and above which they may
be considered to be reliable.

The investigation of the argon trimer revealed that the dissociation process
\cite{Perez10a} discussed earlier with Gaussian approximations
in full detail \cite{Cartarius11a} is correctly described by the frozen
Gaussian imaginary time propagator. It appears in the temperature
range for which the first-order correction is small. Furthermore, the
first-order corrected results are comparable with a fully-coupled thawed
Gaussian investigation of the system, i.e., the addition of the first-order
correction term improves the thermodynamic estimates. The numerical cost of
the first-order corrected frozen Gaussian values is, however, higher than that
of the thawed Gaussian partition function.

The series expansion of the Boltzmann operator exists also for thawed Gaussian
propagators \cite{Shao06a} and has been applied to a one-dimensional system
\cite{Conte10a}. Since the thawed Gaussian propagator approximates the
exact mean energy of the argon trimer at low temperatures better than its
frozen Gaussian counterpart discussed here, it will be interesting to see
whether this can be confirmed with the correction term and whether the
breakdown of the approximation lies at a lower temperature than the frozen
Gaussian results.

Since most effects in rare gas clusters such as structural transformations
or dissociations appear at low temperatures \cite{Neirotti00a,Predescu03a,%
  Frantsuzov04a,Predescu05a,White05a,Frantsuzov06a,Perez10a,Frantsuzov08a}
it will be of value to investigate these properties with the series
expansion. The correction term should be used to verify the validity
of the Gaussian approximations in these cases, in particular, where strong
differences are found between the approximate quantum computations and a
purely classical theory \cite{Frantsuzov06a}.

\section*{Acknowledgements}
  This paper is dedicated to Professor Delgado-Barrio, whose paper on Ar
  clusters was the central impetus for our present investigation.  H.C. is
  grateful for a Minerva fellowship. This work was supported by a grant of the
  Israel Science Foundation.

\appendix

\section{First-order correction term to the partition function for a
  potential expressed in terms of Gaussians}
\label{app:correction_terms}

The correction operator for the frozen Gaussian imaginary time propagator
can be written as \cite{Zhang09a}
\begin{subequations}
  \begin{multline}
    \langle \bm{x}' | C(\tau) | \bm{x} \rangle
    = \langle \bm{x}' | \left ( -\frac{\partial}{\partial \beta} - H \right )
    K_0(\tau) | \bm{x} \rangle
    = \det(\bm{\Gamma}) \\
    \times \exp \Biggl ( -\frac{\hbar^2}{4} \mathrm{Tr}(\bm{\Gamma})
      \tau \Biggr ) \sqrt{\det \left ( 2 \left [ \bm{1} - \exp (-\hbar^2 
          \bm{\Gamma} \tau) \right ]^{-1} \right )} \\
    \times \exp \Biggl ( -\frac{1}{4} [\bm{x}' - \bm{x}]^\mathrm{T} \bm{\Gamma}
      [\tanh(\hbar^2 \bm{\Gamma} \tau/2)]^{-1} [\bm{x}'-\bm{x}] \Biggr ) \\
    \times \int \frac{d\bm{q}^{3N}}{(2\pi)^{3N}}
    \Delta V(\bm{x}',\bm{x},\bm{q}(\tau/2)) \exp \Biggl (-2 \int_0^{\tau/2} d\tau' \langle V(\bm{q}(\tau')) \rangle \\
      - [\bm{\bar{x}}-\bm{q}(\tau/2)]^\mathrm{T} \bm{\Gamma}
      [\bm{\bar{x}}-\bm{q}(\tau/2)] \Biggr ) ,
    \label{eq:corr_FG}
  \end{multline}
  where the energy difference operator is found to be
  \begin{multline}
    \Delta V(\bm{x}',\bm{x},\bm{q}(\tau/2)) = \frac{\hbar^2}{4} \left (
      -\mathrm{Tr}(\bm{\Gamma}) + [\bm{x}' - \bm{x}]^\mathrm{T}
      \frac{\bm{\Gamma}^2}{2} [\bm{x}' - \bm{x}] \right ) \\
    + \langle V(\bm{q}(\tau/2)) \rangle + \frac{\hbar^2}{2} \Biggl ( 
    [\bm{\bar{x}}-\bm{q}(\tau/2)]^\mathrm{T} \bm{\Gamma}^2 [\bm{\bar{x}}
    -\bm{q}(\tau/2)] \\
    + [\bm{x}' - \bm{x}]^\mathrm{T} \bm{\Gamma} \coth \left (
      \frac{\hbar^2 \tau}{2} \bm{\Gamma} \Biggr ) \bm{\Gamma}
      [\bm{\bar{x}}-\bm{q}(\tau/2)] \right ) \\
    + [\bm{\bar{x}}-\bm{q}(\tau/2)] \cdot \langle \nabla V(\bm{q}(\tau/2))
    \rangle - V(\bm{x}') .
    \label{eq:Vdiff}
  \end{multline}
\end{subequations}

Since most parts of the propagator \eqref{eq:prop0} and of the correction
operator \eqref{eq:corr_FG} include only simple Gaussians or polynomials in
$\bm{x}$ and $\bm{x}'$ the matrix element of the product of the zeroth-order
propagator and the correction operator needed for computation of the
first-order correction term may be calculated analytically. Only the parts
including the potential require in general a numerical computation. However,
assuming a Gaussian form \eqref{eq:Gaussian_fit} for the potential enables one
to calculate this part analytically as well.
That is, all $\bm{x}'$ and $\bm{x}$ integrations can be performed analytically
and only the $\bm{q}$ and $\tau$ integrations remain for the numerical
evaluation. After evaluating the $\bm{x}'$ and $\bm{x}$ integrations in
Eq.\ \eqref{eq:pf_corr1} we can write the first-order correction as
\begin{align}
  Z_\mathrm{C1}(\beta) &= \frac{\det(\bm{\Gamma})^2}{(2\pi)^N} \exp \left (
    -\frac{\hbar^2}{4} \mathrm{Tr}(\bm{\Gamma}) \beta \right ) \notag \\ &\quad
  \times \int_0^\beta d\tau \sqrt{\det \left ( \left [ \bm{1} - \exp (-\hbar^2
        \bm{\Gamma} (\beta - \tau)) \right ]^{-1} \right )} \notag \\ &\quad
  \times \sqrt{\det \left ( \left [ \bm{1} - \exp (-\hbar^2 \bm{\Gamma}
        \tau) \right ]^{-1} \right )}  \notag \\ &\quad
  \times \int d\bm{q}_1^{3N} \int d\bm{q}_2^{3N}
  \Delta V_\mathrm{S}(\bm{q}_1,\bm{q}_2,\beta,\tau) \notag \\ &\quad
  \times \exp \left (-2 \int_0^{(\beta-\tau)/2} d\tau' \langle V(\bm{q}_1(\tau'))
    \rangle \right ) \notag \\ &\quad
  \times \exp \left (-2 \int_0^{\beta/2} d\tau' \langle V(\bm{q}_2(\tau'))
    \rangle \right ) ,
\end{align}
where the single trajectory contribution to the energy difference
operator is denoted as
\begin{multline}
  \Delta V_\mathrm{S}(\bm{q}_1,\bm{q}_2,\beta,\tau) =
  \exp \left ( -\frac{1}{2}\Delta \bm{q}(\beta,\tau)^\mathrm{T} \bm{\Gamma}
    \Delta \bm{q}(\beta,\tau) \right ) \\ 
  \times \Biggl \{ \frac{1}{\sqrt{\det( \bm{B}(\beta,\tau)) 
      \det(2\bm{\Gamma})}} \biggl [ \frac{\hbar^2}{4} \Biggl ( - \frac{1}{2} 
    \mathrm{Tr}(\bm{\Gamma}) \\ 
    + \frac{1}{4} \mathrm{Tr}(\bm{B}(\beta,\tau)^{-1} \bm{\Gamma}^2) 
    + \frac{1}{2} \Delta \bm{q}(\beta,\tau)^\mathrm{T} \bm{\Gamma}^2 \Delta \bm{q}(\beta,\tau)
  \Biggr ) \\ + \langle V(\bm{q}_2(\tau/2)) \rangle
  +  \frac{1}{2} \Delta \bm{q}(\beta,\tau)^\mathrm{T} \langle \nabla
  V(\bm{q}_2(\tau/2)) \rangle \biggr ] \\
  - V_\mathrm{SG}(\bm{\bar{q}},\beta,\tau) \Biggr \} ,
  \label{eq:singleqdeltaV}
\end{multline}
the difference trajectory is defined as
\begin{subequations}
  \begin{equation}
    \Delta \bm{q}(\beta,\tau) = \bm{q}_1((\beta-\tau)/2)-\bm{q}_2(\tau/2) ,
  \end{equation}
  and the middle point is
  \begin{equation}
    \bm{\bar{q}}(\beta,\tau) = \frac{1}{2} ( \bm{q}_1((\beta-\tau)/2)
    + \bm{q}_2(\tau/2) ).
  \end{equation}
\end{subequations}
The integrated potential $V_\mathrm{SG}(\bm{\bar{q}},\beta,\tau)$
can be written in a form similar to the Gaussian average of the Gaussian
fitted potential \eqref{eq:Gaussian_fit}, which was introduced by Frantsuzov
et al.\ \cite{Frantsuzov04a} for a fully-correlated multi-dimensional Gaussian
propagator. It consists of a sum of Gaussians,
\begin{multline}
  V_\mathrm{SG}(\bm{\bar{q}},\beta,\tau) = \sqrt{\frac{
      \det(\bm{C}(\beta,\tau))}{\det(\bm{B}(\beta,\tau) + \bm{\Gamma}/2)}} \\
  \times \sum_{j<i} \sum_p c_p \sqrt{\frac{\det(\bm{D}_{ij}(\beta,\tau))}
    {\det(\bm{D}_{ij}(\beta,\tau) + \alpha_p)}} \\ \times
  \exp \left ( - \left [ \bm{\bar{q}}_i(\beta,\tau)
      - \bm{\bar{q}}_j(\beta,\tau) \right ]^\mathrm{T}
    \bm{F}^{(p)}_{ij}(\beta,\tau) \left [ \bm{\bar{q}}_i(\beta,\tau)
      - \bm{\bar{q}}_j(\beta,\tau) \right ] \right ) ,
\end{multline}
with the matrices
\begin{subequations}
  \begin{align}
    \bm{B} (\beta,\tau) &= \frac{1}{4} \bm{\Gamma} \left [ \tanh ( \hbar^2
      \bm{\Gamma} (\beta - \tau)/2 )^{-1} + \tanh ( \hbar^2 \bm{\Gamma} \tau/2
      )^{-1} \right ] \\
    \intertext{and}
    \bm{C}(\beta,\tau) &= \frac{1}{2} \left ( \bm{B}(\beta,\tau)
      + \frac{\bm{\Gamma}}{2} \right ) \left [ \bm{B}(\beta,\tau) \bm{\Gamma}
    \right ]^{-1}  .
  \end{align}
\end{subequations}
The submatrices $\bm{C}_{ij}(\beta,\tau)$ for the particles $i$ and $j$ of
$\bm{C}(\beta,\tau)$ are required for
\begin{subequations}
  \begin{align}
    \bm{D}_{ij}(\beta,\tau) &= \left ( \bm{C}_{ii}(\beta,\tau)
      + \bm{C}_{jj}(\beta,\tau) - \bm{C}_{ij}(\beta,\tau)
      - \bm{C}_{ji}(\beta,\tau) \right )^{-1} \\
    \intertext{and}
    \bm{F}^{(p)}_{i,j}(\beta,\tau) &= \alpha_p -\alpha_p^2 \left ( \alpha_p
      + \bm{D}_{ij}(\beta,\tau) \right )^{-1} .
  \end{align}
\end{subequations}

In Eq.\ \eqref{eq:singleqdeltaV} one can clearly see that due to the Gaussian
in $\Delta \bm{q}$ the contribution of a pair of trajectories $\bm{q}_1(\tau)$
and $\bm{q}_2(\tau)$ will only be nonvanishing if the distance between them
is small. This implies that the restriction of $\bm{q}_1$ to the confining
sphere defined by $R_\mathrm{c}$ will automatically impose a restriction on
$\bm{q}_2$.

\end{document}